# First Observation of Electron Transfer Mediated Decay in Aqueous Solutions: A Novel Probe of Ion Pairing

Isaak Unger,[1] Robert Seidel,[1] Stephan Thürmer,[1,‡] Emad F. Aziz,[1,2] Lorenz S. Cederbaum,[3] Eva Muchová,[4] Petr Slavíček,[4,*] Bernd Winter,[1,*] and Nikolai V. Kryzhevoi[3,*]

[1] *Helmholtz-Zentrum Berlin für Materialien and Energie, Methods for Material Development, Albert-Einstein-Strasse 15, D-12489 Berlin, Germany*

[2] *Department of Physics, Freie Universität Berlin, Arnimallee 14, D-14159 Berlin, Germany*

[3] *Theoretische Chemie, Physikalisch-Chemisches Institut, Universität Heidelberg, Im Neuenheimer Feld 229, D-69120 Heidelberg, Germany*

[4] *Department of Physical Chemistry, University of Chemistry and Technology, Technická 5, 16628 Prague, Czech Republic*

‡Present Address: Department of Chemistry, Kyoto University, Kitashirakawa-Oiwakecho, Sakyo-Ku, Kyoto 606-8502, Japan

*Corresponding authors: nikolai.kryzhevoi@pci.uni-heidelberg.de, bernd.winter@helmholtz-berlin.de, petr.slavicek@vscht.cz

A major goal of many spectroscopic techniques is to provide comprehensive information on the local chemical environment. Electron transfer mediated decay (ETMD) is a sensitive probe of the environment since it is actively involved in this non-local radiationless decay process through electron and energy transfer steps. We report the first experimental observation of ETMD in the liquid phase. Using liquid-jet X-ray photoelectron spectroscopy we explore LiCl aqueous solution, and detect low-energy electrons unambiguously emerging from the ETMD processes of core-ionized $Li^+$. We interpret the experimental results with molecular dynamics and high-level *ab initio* calculations. By considering various solvation-structure models we show that both water molecules and $Cl^-$ anions can participate in ETMD, with each process having its characteristic spectral fingerprint. Different ion associations lead to different spectral shapes. The potential application of the unique sensitivity of the ETMD spectroscopy to the local hydration structure and ion pairing is discussed.

Site-selectivity and sensitivity to the local chemical environment have made X-ray photoelectron spectroscopy one of the powerful tools for probing both gas-phase and condensed matter. The deep inner-shell electron hole, created by photoionization, is followed by relaxation processes which provide additional important insight into electronic structure and correlation in the valence-electron region. One such process is Auger-electron decay, occurring locally at the site of the initial ionization. Auger spectroscopy has found widespread applications in many areas of research, especially in materials science, surface-composition analysis, and medicine.

In the past several years experimental and theoretical works have demonstrated that also non-local electronic relaxation processes occur, and efficiently compete with the local Auger decay. The best studied process is the intermolecular Coulombic decay (ICD)[1] occurring in weakly interacting systems such as rare gases and hydrogen-bonded complexes[2,3]. In the ICD process, the energy gained after refilling the initial hole created by ionization or excitation is used to eject an electron from a neighboring atom or molecular entity, resulting in the formation of two singly charged units which subsequently separate by Coulomb repulsion. The competition of non-local and local relaxation processes in aqueous solution has been recently examined[4–6]. ICD in an aqueous environment is particularly important because of the production of slow electrons and water radical cations[7,8], and all these particles represent potential threads of damage to biological tissue. On the other hand, control of the ICD efficiency, of the emission site and energy of the ICD electrons[9–11] would be very attractive for various reasons, including applications in medical treatment, or chemical reactions at electrode surfaces immersed in solutions, e.g., in material and energy research.

The present study addresses yet a different and more complex non-local electronic relaxation process, electron transfer mediated decay (ETMD)[12], which remains largely unexplored. The first step in ETMD, unlike in Auger decay and ICD, is the refilling of the created vacancy by an electron from a *neighboring* atomic or molecular monomer. The energy released is used to ionize either the same electron-donating monomer (ETMD(2) process) or a third monomer in the vicinity (ETMD(3) process). So far, valence-ionized rare-gas clusters were the only systems where ETMD was observed experimentally[13–15]. These finite-size complexes were also in the focus of most of the theoretical ETMD studies, see, *e.g.*, refs.[16–18]. ETMD is, however, a general phenomenon, and has been predicted theoretically to occur in various environments and follow core ionization as well[6].

In the present work we prove experimentally several theoretical predictions. For the first time ETMD is detected in an aqueous solution. Furthermore, we report on the first unequivocal observation of the ETMD processes following core ionization based on experiment. In our quest for a spectral signature we chose a system where no other non-radiative relaxation but ETMD is allowed, and thus the emitted ETMD electrons can be unambiguously assigned. Aqueous LiCl solution is a particularly suitable candidate. As was pointed out in a theoretical study by Müller and Cederbaum[19], core-ionized $Li^+$(aq) cannot decay electronically via Auger or ICD mechanisms since $Li^+$ has no valence electrons, whereas ETMD is possible and proceeds fast, within 20 fs. Yet the detection of ETMD electrons is experimentally challenging, requiring long acquisition times and stability of the position of the liquid microjet in order to identify the ETMD signal on the large background from secondary (inelastically scattered) electrons forming the low-kinetic energy part of the photoelectron spectrum.

In order to explain more specifically what our study is expected to reveal we illustrate in the energy-level diagrams of Figure 1 the most relevant ETMD(2) and ETMD(3) processes following 1s ionization of $Li^+$(aq); other ETMD processes are not shown in the figure but will be considered in our calculations. The ETMD(2)$_W$ process shown in Figure 1A involves a single water molecule W which donates a valence electron to fill the $Li^{++}$(aq) core hole. The released energy is then used to ionize another valence electron from the same water molecule. In the experiment, the kinetic energy of this latter electron, ejected into vacuum is detected. Contrary to ETMD(2)$_W$, in ETMD(3)$_{W,W}$ (Figure 1B) the energy released in the first electron-transfer step is used to ionize a second water molecule. In the presence of a counter ion, $Cl^-$ in this study, the energy released in the first ETMD step can be also transferred to this anion, causing electron detachment. The respective ETMD(3)$_{W,Cl}$ process is depicted in Figure 1C.

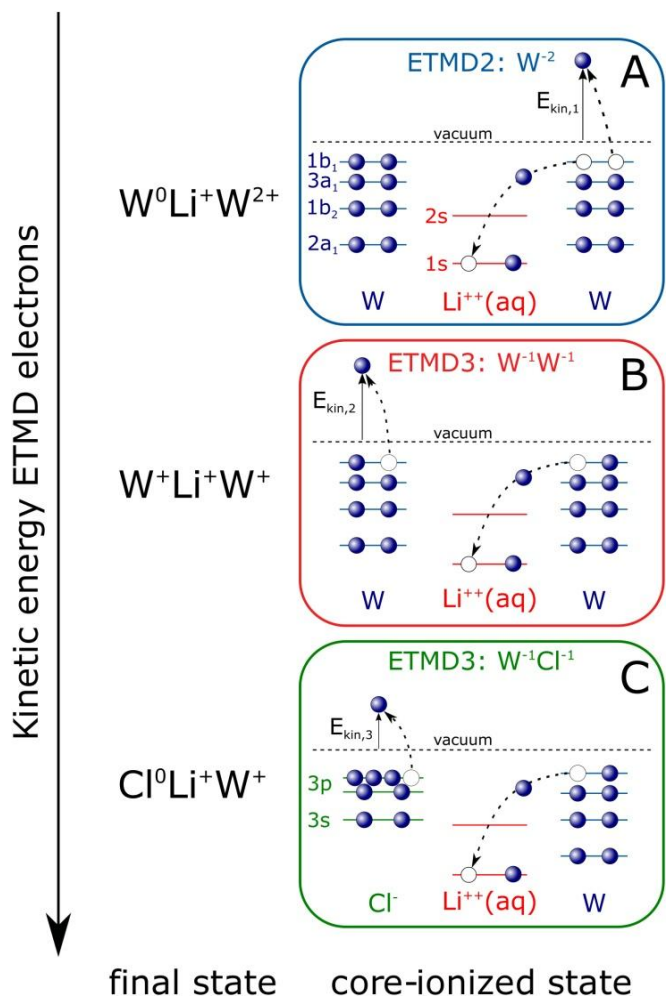

**Figure 1** Depiction of the most relevant ETMD processes, ETMD(2)$_W$ (top), ETMD(3)$_{W,W}$ (center) and ETMD(3)$_{W,Cl}$ (bottom), in LiCl aqueous solution; see text for details. Starting point in each case is the 1s core-level ionization of Li$^+$(aq) forming Li$^{++}$(aq). Subscript W denotes a water molecule. KE denotes the kinetic energies of electrons emitted in ETMD processes (briefly ETMD electrons) which are measured in the experiment. The respective final ETMD states and the relative kinetic energies of the ETMD electrons are shown at the left side.

Apparently, the electrons emitted in different ETMD processes have different kinetic energies thus leading to characteristic spectral shapes depending on the specific local atomic environment. This is indeed confirmed by the experimental spectra which exhibit several features in the energy distribution of ETMD electrons. From a fundamental science point of view this is an important result demonstrating that the ETMD process not only is operative in aqueous solutions but can be used to generate low-energy electrons within a small tunable energy window, independent of the applied ionization energy. And vice versa, one can exploit the structured ETMD spectrum to infer local solvation-structure details, including various forms of anion–cation pairings (solvent-separated ion pairs, contact or solvent-shared ion pairs). The potential of this latter aspect will be considered here in some depth, based on electronic-structure calculations for few simple model configurations of $Li^+$ solvation. Our calculations do not resolve the exact solvation structures of this particular aqueous solute, and are rather meant to demonstrate the potential of this spectroscopy for structure probing.

We would like to point out that ion pairing in solution, including LiCl aqueous solutions, has been intensively studied using a wide range of experimental techniques, such as conductivity measurements, potentiometry, linear UV absorption spectroscopy, NMR, Raman spectroscopy, mass spectrometry, THz absorption, femtosecond mid-infrared, and dielectric relaxation spectroscopies.[20–23] The latter methods are typically based on the real-time monitoring of the ultrafast reorientation dynamics of water molecules, performing femtosecond laser pump-probe experiment. THz absorption spectroscopy addresses the ion–water vibrations, and the dependence of the linear absorption on concentration reveals insight into specific ion-pairing situations. Ion pairing in solution has been also studied exploiting the advances in laser and X-ray spectroscopies. X-ray scattering provides structural details of ion pairing, inferred from the distribution of distances between unlike ions and between ions and water molecules [e.g., refs. [24–26]]. Studies of LiCl aqueous solutions up to 4 molal concentration, combining X-ray scattering and MD simulations, have provided insight into local solution structure, manifested by a decrease of hydrogen bonding with respect to neat water, which is accompanied by an increase of contact ion pairs and a decrease of solvent-separated ion pairs[27]. Ion pairing was studied also by X-ray absorption spectroscopy which is arguably not an as powerful method for studying ion pairing as some of the aforementioned techniques[28].

## Results and Discussion

ETMD spectra from 3 and 4.5 M LiCl aqueous solutions, measured at 180 and 200 eV photon energies, respectively, are shown in tiers (b) and (c) of Figure 2. The photon energies applied are well above the Li$^+$(aq) ionization threshold (60.4 eV[29]) in water. Energies of the measured electrons are presented as kinetic energies (top scale) and as double-ionization energies of the final states (bottom scale). We refer to the latter scale in the following discussion. Spectra (b) and (c) are obtained from subtraction of a neat-water photoelectron spectrum from the respective solution spectrum. As an example we show in tier (a) the data for 3 M concentration, displayed over the range of relevant electron kinetic energies. With respect to the neat-water spectrum (blue) a small signal increase near 30 eV kinetic energy, which will be assigned to ETMD electrons, is observed in the solution spectrum (red). Note that both spectra (a) are dominated by the emission of inelastically scattered (photo)electrons which give rise to the characteristic rather structure-less large electron signal steadily increasing towards lower kinetic energies. Since the ETMD signal of interest is much smaller than the background signal from inelastic electrons the differential spectra (b) and (c) exhibit a rather large signal-to-noise ratio. In order to show that the data are yet statistically significant the as-measured individual data points have been binned and the resulting error bars have been determined. Results for 5-point binning are presented by the black full circles, and additional smoothing yields the green line. As detailed in the experimental section better statistics of the ETMD spectra cannot be achieved with the present setup using standard electron-energy detection schemes. Very likely, future electron-electron-coincidence measurements will deliver higher-quality spectra. Such an experimental approach has been recently demonstrated for the inner-valence ionization of water clusters, where ICD electrons could be distinguished from the direct photoelectrons[8].

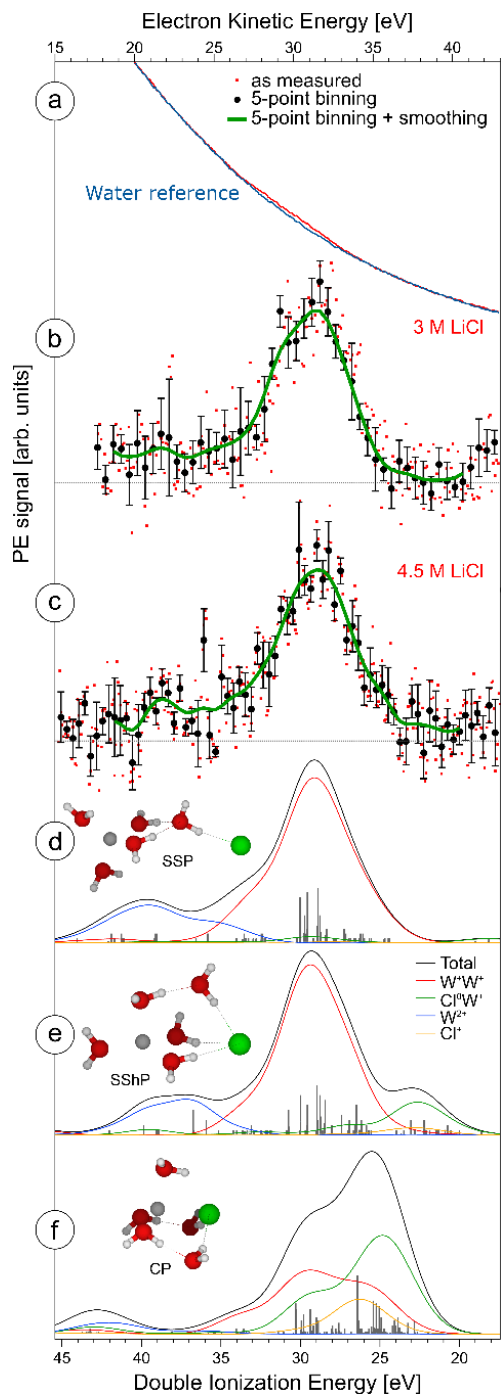

**Figure 2** ETMD spectra from LiCl aqueous solutions shown on the kinetic energy (top) and double-ionization energy (bottom) scale. (a) and (b): Experimental ETMD spectra from 3.0M concentration resulting from core ionization of Li+(aq) at 180 eV photon energy. In (a) we show the as-measured spectrum (in red), including the reference spectrum of neat water (which only contains contributions from inelastically scattered photoelectrons; in blue). Tier (b) is the resulting difference spectrum, solution

minus water, yielding the red dots. Black dots result from 5-point-binning of the red dots, and the green line results from additional smoothing. Tier (c) shows the analogous data as in tier (b) but for 4.5M concentration, and the photon energy was 175 eV. (d)-(f): Theoretical ETMD spectra (black solid curves) computed for the SSP, SShP, and CP cluster models, respectively; see text for explanations. Energies and intensities of individual transitions are shown also as sticks. Each stick has been convoluted by a Gaussian with fwhm of 3.6 eV. The geometries of the cluster models are depicted in the insets (red: oxygen; green: $Cl^-$; grey: $Li^+$; white: hydrogen). The theoretical ETMD spectra are decomposed into various contributions corresponding to different ETMD processes (color solid curves, see the legend).

It is important to note that the spectra from both solutions are rather similar which leads us to draw two conclusions. First, small concentration dependence between 3-4.5 M suggests that the hydration structure and ion pairing are similar in this concentration range. Second, the spectra are independent of photon energy (at least within the 180-200 eV range studied) which shows that the signal indeed arises from electronic decay, and contributions from direct ionization can be ruled out. We observe a broad structure in the 45-20 eV range, with a dominant peak at 28.5 eV. This peak is attributed to the ETMD(3)$_{W,W}$ processes producing two outer-valence ionized water molecules $H_2O^+(3a_1)$. This can be qualitatively seen from the consideration of two electrostatically interacting water cations. In the case of perfect electronic screening the corresponding ETMD signal would appear at 27 eV (2×13.50, where 13.50 eV is the binding energy of the $3a_1$ orbital in liquid water[29]) that is close to the experimental value. The small energy difference (~1.5 eV) is likely due to residual repulsion energy which cannot be screened completely in aqueous media. The theoretical calculations described below provide additional support to this assignment.

Aqueous structure of LiCl solution has been extensively studied experimentally and by means of molecular dynamics simulations[24,30–35], and its structure, in particular ion pairing, is rather well understood. We apply here the classical molecular dynamics simulations with electronic continuum correction to investigate the ion pairing. Below we show that the solvent-shared arrangement is expected to prevail in the LiCl solutions in the concentration range studied in our work. Figure 3 shows the radial distribution functions for $Li^+$—$Cl^-$ and $Li^+$—O for 3M, 4.5M and 6M aqueous solutions. The 6M case was considered to gain insight into the structure of the LiCl aqueous solution with higher salt concentration than in the experiment. While the Li$^+$―O curves are nearly indistinguishable, the radial distribution function for Li$^+$―Cl$^-$ varies with concentration. The same concentration dependences have been shown by Nasr. et al.[24] for salt concentrations in the range 3 to 6m. This behavior is well seen for the first maximum at $r$ = 2.36 Å corresponding to a contact ion pair arrangement. The second peak of this curve at $r$ = 4.62 Å attributed to a solvent-shared ion pair structure reveals however only minor dependence on concentration. A histogram showing the minimum distance between a selected Li$^+$ cation and the closest Cl$^-$ anion for all three concentrations is depicted in Supplementary Figure S1A. Quantitatively, the fraction of the contact ion pairs is 17 % for the 3M, 27 % for the 4.5M and 40 % for the 6M solution (see Supplementary Figure S1B). The by far dominant structure is thus the solvent-shared ion pair structure with Li$^+$ and Cl$^-$ being about 4.5 Å apart; yet, for the 3M solution, there is still a chance for a solvent-separated ion pair. Another way to quantify liquid structure is through coordination numbers. Our results tabulated in Supplementary Table S1 agree reasonably well with other simulations[34]. As one can see, the chloride anion occupying the first solvation shell becomes a more prominent structure with increasing concentration; yet contact ion pairs are clearly not dominating even for the highest concentration. Another type of structural analysis based on permutation invariant vector clustering that reveals the most populated structural motives is presented in Supplementary Information in Figure S2.

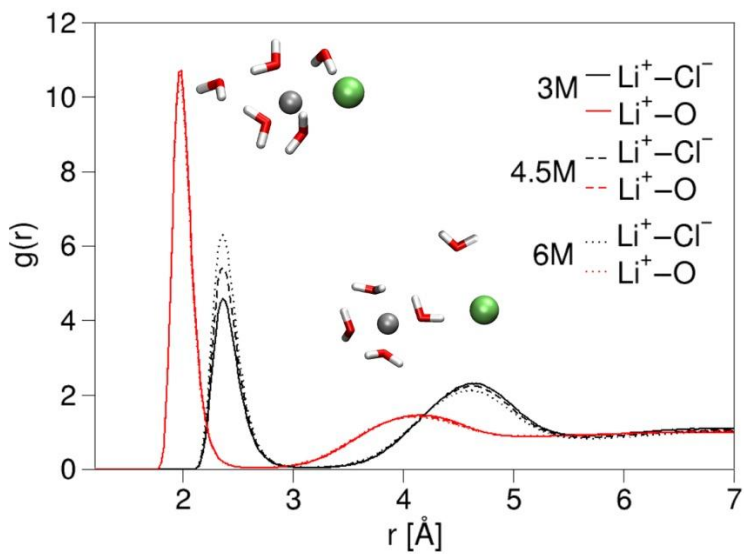

**Figure 3** Radial distribution functions for 3M, 4.5M and 6M LiCl aqueous solutions. Snapshots from molecular dynamics simulations corresponding to a contact ion pair (left) and a solvent-shared ion pair (right) are shown in insets (grey: Li$^+$; green: Cl$^-$).

Next, we discuss the ETMD spectra. We consider three cluster models: SSP for solvent-separated ion pairs, SShP for solvent-shared ion pairs and CP for contact ion pairs. These models are described in the computational methods. The corresponding spectra are shown in tiers (d)-(f) of Figure 2, respectively. The spectrum of the SShP model is expected to fit best to the experimental data due to prevalence of solvent-shared arrangements in solution. Since the calculations were performed for the gas phase, the spectra need to be shifted in energy to account for solvent effects. The dominant contribution to this shift results from the long-range polarization which acts differently on different ETMD states. Note that in the photoemission spectra of aqueous electrolytes, the energies of the photoelectron peaks of both the solvent water molecules and the solutes are virtually unaffected by the electrolyte concentration[29] (this is not the case in finite-size systems where the peak positions depend strongly on the local solvation structure, including ion-pairing). Apparently, water is capable to screen very efficiently the electrostatic interactions between neighboring molecules. Thus, we can assume that the energy position of the same ETMD state in the experimental spectra of aqueous solutions does not depend on the environment. In particular, it is not important for the double-ionization energy of two neighboring water units involved in ETMD(3)$_{W,W}$ whether their environment contains water molecules or ions. Since solvent-shared ion pairs prevail in the measured LiCl aqueous solutions, and the ETMD(3)$_{W,W}$ processes play the most important role in such structural units (see below), we chose the ETMD(3)$_{W,W}$ states for the alignment of all spectra. Accordingly, the spectra of the SSP, SShP and CP models were shifted to lower double-ionization energies by 7.65 eV, 5.03 eV and 3.89 eV, respectively. It should be noted that in contrast to the SSP and SShP models, the ETMD(3)$_{W,W}$ processes in the CP model mostly contribute to the high-energy shoulder and not to the main peak which originates from different processes. Therefore, the main peak in the theoretical spectrum of this model does not coincide in energy with the main peaks in the experimental spectra.

We now discuss the spectral shapes in more detail. The ETMD spectrum of the SShP model exhibits a well-defined main peak at 28.5 eV and two smaller peaks, one spreading between 35 and 40 eV and another one at 22 eV. The ETMD spectrum of the SSP model has a similar structure except for the missing peak at the low double-ionization energy side, and a more pronounced shoulder of the main peak at 33 eV. The decomposition of each spectrum into various contributions reveals that the main peak arises essentially from ETMD(3)$_{w,w}$ processes. Interestingly, only a small fraction of the water cations created in these processes is found in the cationic ground state, *i.e.*, with the 1b$_1$ electron removed. The reason will be discussed below. Most of the water molecules eject electrons from the deeper-lying orbitals during electronic decay. According to our calculations, the main peak mostly comprises the ETMD(3)$_{3a1,3a1}$ states where two 3a$_1$ vacancies are produced, each on a different water molecule. Its low-energy part at approximately 26 eV is attributed to the ETMD(3)$_{3a1,1b1}$ processes creating pairs of 1b$_1$- and 3a$_1$-ionized water molecules while the high-energy part at 33 eV corresponds to the processes creating pairs of 1b$_2$- and 3a$_1$-ionized water molecules. Both spectral regions are extremely sensitive to the orientations of water molecules in the first solvation shell of the metal ion as seen from Supplementary Figure S3.

The spectral region near 33 eV contains also some contributions from ETMD(2)$_w$ processes. The main spectral domain of these processes lies however at higher energy, and coincides with the peak which spreads from 35 to 40 eV. As for ETMD(3)$_{w,w}$, also ETMD(2)$_w$ mostly involves 3a$_1$ electrons of water molecules, and water dications with two vacancies in the 3a$_1$ orbital are the main products of this decay channel. The high probability of the 3a$_1$ electrons to participate in ETMD in the SShP and SSP models results from favorable orientations of the water monomers in the first solvation shell of Li$^+$. Their oxygen atoms point toward the cation which maximizes the overlap of the 3a$_1$ orbitals with the 1s orbital of Li$^+$. Although efficiencies of the individual ETMD(3)$_{w,w}$ and ETMD(2)$_w$ processes may be comparable, the ETMD(3)$_{w,w}$ peak acquires more intensity. This is because the total ETMD(2)$_w$ efficiency is approximately proportional to the number of water monomers in the first solvation shell (other water molecules are far less prone to ETMD(2)), whereas the total ETMD(3)$_{w,w}$ efficiency correlates with a much larger number of water pairs, predominantly with one or two water monomers from the first solvation shell. It is also worth mentioning that the relative intensities of the ETMD(2)$_w$ and ETMD(3)$_{w,w}$ signals are very similar in tiers (d) and (e) which is attributed to similar structures of the first solvation shells of Li$^+$ in the respective cluster models.

Although the arrangements of water molecules in the immediate neighborhood of $Li^+$ in the SSP and SShP cluster models are very similar, structures differ significantly beyond the first solvation shell of the cation. In the SSP model the chloride anion is at far distance from the metal, appearing only in the third solvation shell. The two counter ions are much closer to each other in the SShP configuration, wherein they are separated by only two bridging water molecules. These structural differences are reflected in the ETMD spectra, especially in their low-energy parts where $ETMD(3)_{W,Cl}$ processes contribute. As can be inferred from Figure 2, the $ETMD(3)_{W,Cl}$ efficiency depends strongly on the counter-ion separation. In the SShP configuration, the $ETMD(3)_{W,Cl}$ peak at 22 eV is well resolved despite the fact that $Li^+$ and $Cl^-$ are separated by one solvation shell. In the SSP model, this peak is nearly absent.

The ETMD spectrum changes drastically when one of the water molecules nearest to $Li^+$ is substituted by $Cl^-$, which leads to a contact ion pair. First, the efficiency of $ETMD(3)_{W,Cl}$ increases substantially. The $ETMD(3)_{W,Cl}$ signal, which appears only as a secondary peak in the spectrum of the SShP configuration, now acquires high intensity and becomes the main peak, at 24 eV, as seen in tier (f) of Figure 2. It should be noted that aside from the $ETMD(3)_{W,Cl}$ contribution, this peak contains also large contributions from the $ETMD(3)_{W,W}$ and $ETMD(2)_{Cl}$ processes. The latter produce $Cl^+$ cations and seem to appear only in the CP model. Since $Li^+$ now has less water molecules in the first solvation shell, the impact of the $ETMD(3)_{W,W}$ and $ETMD(2)_W$ processes on the total ETMD spectrum decreases. The formation of a contact ion pair also modifies the orientation of water molecules such that the 1s orbital of $Li^+$ now better overlaps with the $1b_1$ orbitals of water, and the efficiency of the $ETMD(3)_{3a1,1b1}$ processes producing pairs of $1b_1$- and $3a_1$-ionized water molecules increases. At the same time fewer $3a_1$-ionized water pairs are produced by $ETMD(3)_{3a1,3a1}$. These trends can be recognized from the red curve in tier (f) of Figure 2 whose right and middle peaks originate from the $ETMD(3)_{3a1,1b1}$ and $ETMD(3)_{3a1,3a1}$ processes, respectively; compare with the red curves in tiers (d) and (e).

As seen from Figure 2 (tiers (b) and (c)), the agreement between the experiment and the theory is very good. The main peak with its shoulder originating from the $ETMD(3)_{W,W}$ processes, and the $ETMD(2)_W$ peak emerging at a higher double-ionization energy are well reproduced by the theory. The situation with the $ETMD(3)_{W,Cl}$ peak is less clear. By taking into account the above consideration that a certain ETMD state should have the same energy irrespective of the particular environment, and assuming that ETMD(3)$_{W,Cl}$ mostly produces $3a_1$-ionized water molecules and neutral chlorines (as predicted by the theory), the corresponding peak should appear at 23.1 eV (13.5 eV+9.6 eV, where 9.6 eV is the detachment energy of Cl$^-$(aq)). The spectrum of the 4.5M LiCl solution exhibits a vaguely visible structure at 22 eV. No signal at this energy is, however, seen in the spectrum of the 3M solution which might indicate at a larger fraction of solvent-separated ion pairs at this concentration. One needs to perform additional measurements with a higher signal-to-noise ratio in order to prove the relation of this signal with the ETMD(3)$_{W,Cl}$ processes and its dependence on salt concentration.

## Conclusions

We have reported on the first experimental observation of ETMD processes in aqueous media. The emergence of the ETMD process has been unambiguously proven by measuring secondary electrons in LiCl aqueous solution upon core ionization of the Li$^+$(aq) cation where all other non-radiative relaxation channels are energetically closed. The ETMD process was so far identified exclusively in finite rare-gas clusters, and this work has thus settled ETMD as a general phenomenon. Additionally, the present study has demonstrated experimentally for the first time the feasibility of ETMD processes following core ionization. Note at this point that the ETMD signal will be particularly strong when other non-radiative channels such as Auger decay are closed. Such a situation occurs for a number of ions upon of formation of valence-ionized states.

Identification of ETMD in aqueous solutions is relevant from a perspective of radiation chemistry. Similarly as for the better known ICD process, slow electrons and water radical cations are produced within this decay. As both of the above species are highly reactive, the ETMD pathway should also be considered in modeling radiation chemistry processes. Furthermore, similarly to ICD, the occurrence of the ETMD process can be envisioned as a unique means to generate low-energy electrons, initiating chemical reactions, e.g., at biological surfaces or at the electrode–solution interfaces as encountered in many areas of material-energy research. Clearly, the present work on Li$^+$(aq) is just a very first step into a new research field, and even the unequivocal demonstration of ETMD in more complex systems is experimentally challenging.

It is tempting to ask whether the spectroscopy of ETMD electrons can be transformed into a novel spectroscopy tool for liquid-phase investigations. Our theoretical computations have revealed that ETMD spectra are sensitive to the structure of the first solvation shell around the initially ionized lithium ion, reflecting orientations of solvent water molecules, ion-water distances (see Supplementary Figures S3 and S4) and ion pairing. Different features are clearly visible in the theoretical ETMD spectra obtained for different cluster models which are representative of the distinctive ion pairing situations present in aqueous solutions. The experimentally measured spectra are in agreement with the known ion associations in concentrated LiCl aqueous solutions. The present experiment is however limited in its sensitivity due to a relatively small signal-to-noise ratio and the restricted range of salt concentrations. The quantitative analysis of ion pairing based on recording ETMD electrons is therefore not possible at the moment but the present work justifies that this analysis can be done in future studies. The ETMD signal can be improved, e.g., by using a combination of coincidence technique and a magnetic bottle electron analyzer. ETMD spectroscopy may then become a powerful tool for studying various properties of aqueous solutions. ETMD may find applications also in systems with organic or hybrid solvents, e.g., in Li-ion or metal-air batteries where the knowledge of ion pairing and local solvation structure is essential[36].

The experimental spectra are presently interpreted with the help of *ab initio* and molecular dynamics simulations. More experimental data would be required, including those for different aqueous solutions, to find out whether ETMD spectroscopy can identify ion pairing solely based on the experimentally measured spectra. Our present results indicate that the position of the ETMD spectra in the liquid phase can be reasonably well estimated from the binding energies of participating electron measured by direct photoemission. Further experiments and theoretical calculations are needed to firmly establish this relation. It would be furthermore interesting to find out whether the ETMD spectra reflect the structural changes connected, e.g., to changing temperature or phase transitions.

With respect to other techniques for probing local structure, and particularly ion pairing, ETMD appears to be a complementary tool for structure analysis. Unlike e.g., the non-linear femtosecond infrared or the dielectric relaxation spectroscopies, where vibrational dynamics of certain modes is tracked, ETMD relies on a completely different interaction and relaxation process. Here, the immediate neighbors are probed, with large sensitivity to distance and charge state, through an electron-transfer process between molecules or atoms. At the same time the outcome of ETMD, i.e., the creation of a very distinct doubly-ionized molecular structure, comprising two reactive molecular entities, offers a unique possibility to initiate chemical reactions, for instance at the solid–solution of protein–solution interface. We currently know very little about such ETMD-initiated chemical reactions, and it is yet to be investigated how to apply this new spectroscopy for *practical* purpose.

## Methods

### Experiment

Autoionization electron spectra from 3 and 4.5 M LiCl aqueous solution were measured from a 15-μm vacuum liquid-water jet, and ionization photon energies were 180 and 175 eV, respectively. Experiments were conducted at the U41-PGM undulator beamline of BESSY II, Berlin. The jet velocity was approximately 80 ms$^{-1}$, and the jet temperature was 6 °C, similar to our previous studies[37]. Electrons were detected with a hemispherical electron analyzer, separated by a 100 μm diameter orifice from the liquid jet at a distance of approximately 300 μm. The two solutions were measured at different times, using different detection geometries. For the 3 M LiCl solution the detection direction was normal with respect to the light polarization vector, whereas for the 4.5 M concentration measurements were performed at the magic angle, at approximately 54.7°. Spectra presented in this work were collected over a total time of 120 min, which has been broken down into two 60-min data collection periods. This comprises equally long measurements of the solution spectra and of neat-water reference spectra. Longer acquisition times have been attempted. However, slight changes of the liquid-jet position with respect to both the photon beam and the electron detector lead to considerable differences in the shape of the distribution of the secondary electrons which made a meaningful subtraction of pair-wise measured neat-water and solution-spectra impossible. The energy resolution of the U41 beamline was better than 200 meV at the incident photon energies used here, and the resolution of the hemispherical energy analyzer is constant with kinetic energy (about 200 meV, at 20 eV pass energy). A small X-ray focal size, 23 × 12 μm$^2$, assures that the gas-phase signal amounts to less than 5% of the total (photo)electron signal. Solutions were prepared by dissolving LiCl (Sigma Aldrich) in highly demineralized water (conductivity ~0.2 micro Siemens/cm).

## Computations

*Molecular Dynamics Simulations*

We modeled the LiCl solution using classical molecular dynamics simulations, assuming 3M, 4.5M and 6M LiCl aqueous solutions. The simulation box for the 3M solution contained 72 Cl$^-$ and 72 Li$^+$ ions and 1259 water molecules in a cubic box with a length of 34.157 Å. The more concentrated systems contained in the same box size 108 Cl$^-$, 108 Li$^+$ ions and 1235 water molecules for the 4.5M, and 144 Cl$^-$, 144 Li$^+$ ions and 1211 water molecules for the 6M solution. The force field parameters were taken from ref.[33], the parameters for Li$^+$ were σ = 1.80 Å and ε = 0.07647 kJ/mol. In all simulations, the rigid SPC/E (extended simple point charge model) water model was used[38]. For lithium and chlorine ions, the electronic continuum correction approach[39] was used, yielding scaled charges of +0.75e and −0.75e. This approach aims to mimic the effect of electronic polarization in an efficient way; it has been used successfully before for LiCl solutions[32]. The simulations runs were 30 ns long using a time step of 1 fs. Simulations were performed at a constant volume and temperature of 300 K maintained by a CSVR thermostat[40] with a time constant of 0.5 ps. Periodic boundary conditions were employed with short-range electrostatic and van der Waals interactions truncated at 1.2 nm and the long-range electrostatic interactions treated by the particle mesh Ewald method[41]. All simulations were performed with the GROMACS 4.5.3 code[42].

*Cluster models and geometry optimization*

We simulated ETMD spectra for three cluster models, representative of the different ion pairing situations occurring in aqueous solutions: contact, solvent-shared and solvent-separated ion pairs. All clusters consist of one Li$^+$ cation, one Cl$^-$ anion, and five solvent water molecules, but differ in their structural arrangement. Although our cluster models are small, which was necessary in order to make *ab initio* computations of the ETMD spectra feasible, they nevertheless capture the essential characteristics of the ETMD processes in aqueous solution. Note that ETMD is a charge-transfer process and therefore involves predominantly the nearest neighbors. The Li$^+$ cation is fully solvated in all clusters, being surrounded by four water molecules in the first solvation shell in the SShP and SSP configurations. In the CP model, one of the nearest water molecules is substituted by the Cl$^-$ anion. The Li$^+$−Cl$^-$ distance was set at 2.4 Å in the CP model, 4.3 Å in the SShP model and 6.4 Å in the SSP model according to the peak positions in the experimental and theoretical radial distribution functions of the LiCl aqueous solution. We also fixed the distances between the ions and the nearest water molecules at the values found in the aqueous LiCl solution, namely $d$(Li$^+$−O) = 1.95 Å and $d$(Cl$^-$−O) = 3.18 Å[24,30–35]. With the above constraints, geometry optimization was then performed for each cluster using the second-order Møller-Plesset perturbation theory in conjunction with the 6-311++G(2d,2p) basis set.

*Simulations of the ETMD spectra*

The double-ionization energies of the clusters were calculated using the second-order algebraic diagrammatic construction method, ADC(2), which is an approximation scheme for the two-particle propagator[43,44]. In these calculations we employed the Dunning's double-ζ DZP basis sets[45] for Li and Cl. The basis set on Cl was additionally augmented with one s-type and one p-type diffuse functions. Water molecules were described in the same way as in our recent work[6], *i.e.*, using the cc-pVDZ basis set[46] for hydrogens and a relativistic pseudopotential basis set for oxygens. The latter was augmented with diffuse (one s-type and one p-type) and polarization (one d-type) functions.

The character of each final dicationic ETMD state was determined with the two-hole population analysis[47]. Within this method, the pole strengths of the computed ADC(2) states are decomposed into contributions originating from configurations with different distributions of the two final outer-valence holes in the system. As these holes may be located either on two different atoms or on the same atom, one distinguishes between "two-site" and "one-site" contributions. The one-site contributions are typically used as relative intensities in the simulated decay spectra[6]. In particular, for constructing the ETMD spectra resulting from 1s ionization of $Li^+$, we selected the one-site Li contributions in the pole strengths. Finally, for comparison with experiment, each spectral line has been convoluted with a Gaussian of full width at half maximum, fwhm=3.6 eV. This broadening accounts for both the effect of vibrational delocalization occurring during the ETMD processes and for the different solvent configurations present in aqueous solutions.

## Acknowledgments

E.M. and P.S. thank Czech Science foundation for a support (project number 13-34168S). NVK and LSC gratefully acknowledge the financial support by the Deutsche Forschungsgemeinschaft. IU and BW also gratefully acknowledge support from the Deutsche Forschungsgemeinschaft (DFG Research Unit FOR 1789). We thank Eva Pluhařová for valuable discussions.

## Author contributions

I.U, R.S., S.T. and B.W. conceived, designed and performed the experiments, and analyzed the experimental data. E.M. and P.S. performed and analyzed the molecular dynamics simulations. N.V.K. computed the theoretical ETMD spectra and analyzed them. B.W, P.S. and N.V.K. co-wrote the paper. All authors discussed the results and commented on the manuscript.

## Competing financial interests

The authors declare no competing financial interests.

# Supplementary Information

# First Observation of Electron Transfer Mediated Decay in Aqueous Solutions: A Novel Probe of Ion Pairing

Isaak Unger,[1] Robert Seidel,[1] Stephan Thürmer,[1,‡] Emad F. Aziz,[1,2] Lorenz S. Cederbaum,[3] Eva Muchová,[4] Petr Slavíček,[4,*] Bernd Winter,[1,*] and Nikolai V. Kryzhevoi[3,*]

[1] *Helmholtz-Zentrum Berlin für Materialien and Energie, Methods for Material Development, Albert-Einstein-Strasse 15, D-12489 Berlin, Germany*

[2] *Department of Physics, Freie Universität Berlin, Arnimallee 14, D-14159 Berlin, Germany*

[3] *Theoretische Chemie, Physikalisch-Chemisches Institut, Universität Heidelberg, Im Neuenheimer Feld 229, D-69120 Heidelberg, Germany*

[4] *Department of Physical Chemistry, University of Chemistry and Technology, Technická 5, 16628 Prague, Czech Republic*

‡Present Address: Department of Chemistry, Kyoto University, Kitashirakawa-Oiwakecho, Sakyo-Ku, Kyoto 606-8502, Japan

*Corresponding authors: nikolai.kryzhevoi@pci.uni-heidelberg.de, bernd.winter@helmholtz-berlin.de, petr.slavicek@vscht.cz

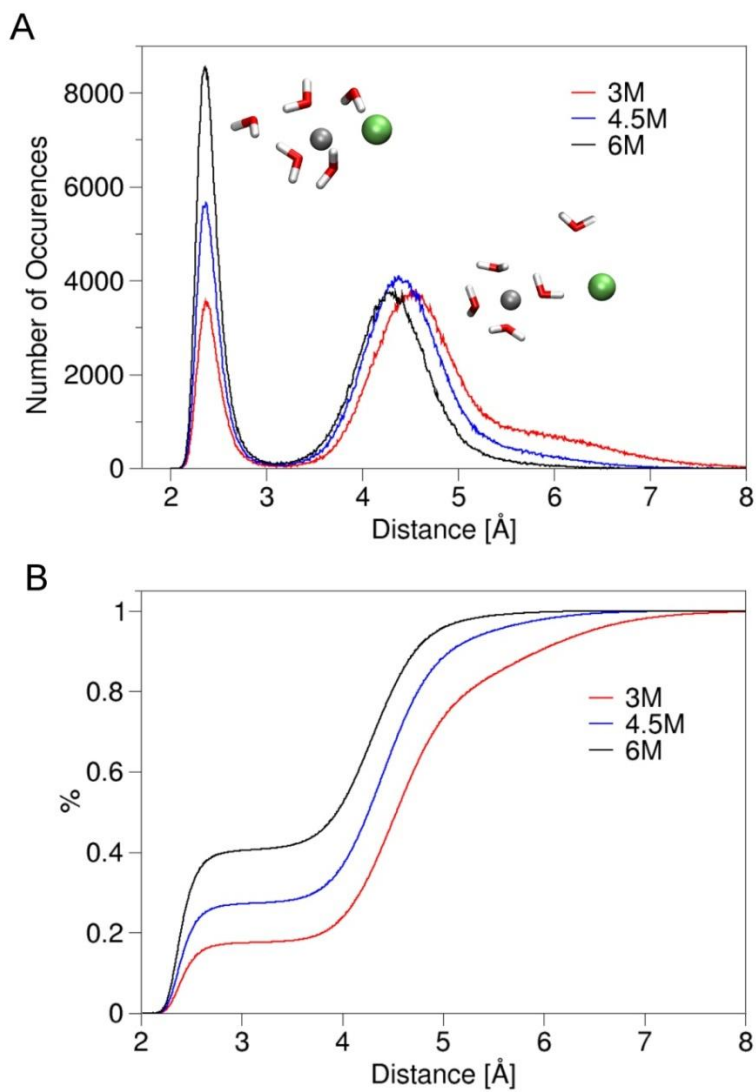

**Supplementary Figure S1 A**: Histogram showing the minimum distance between one selected Li$^+$ ion and the closest Cl$^-$ for the 3M, 4.5M and 6M aqueous solutions of LiCl. **B**: Integration of the curves shown in panel A up to a certain distance value. The first plateau gives the percentage of contact ion pairs: 17% for 3M, 27 % for 4.5M and 40% for 6M solution.
2

**Supplementary Table S1:** Mean coordination numbers of Li$^+$ ion for LiCl solutions of different concentrations. The coordination numbers were calculated as $n_{Li^+}^{Cl^-} = 4\pi\rho_{Cl^-} \int_0^{r_1} g_{LiCl}(r) r^2 dr$, where $\rho_{Cl^-}$ is the Cl$^-$ number density, $g_{LiCl}$ is the radial distribution function of Li$^+$—Cl$^-$ with the first minimum at $r_1$, and as $n_{Li^+}^{H2O} = 4\pi\rho_{H2O} \int_0^{r_1} g_{LiO}(r) r^2 dr$, where $\rho_{H2O}$ is the H$_2$O number density, $g_{LiO}$ is the radial distribution function of Li$^+$—O with the first minimum at $r_1$.

| concentration | coordination number Cl$^-$ | coordination number water |
|---|---|---|
| 3M | 0.180 | 3.816 |
| 4.5M | 0.317 | 3.691 |
| 6M | 0.485 | 3.532 |

**Structural cluster analysis**

In order to reveal the structural motives present in liquids we performed the structural cluster analysis based on the so called permutation invariant vector (PIV)[1] approach on the geometries obtained from our MD simulations. The PIV method partitions the geometries into a few structural clusters (i.e., sets of geometries that are structurally similar). Each one has a cluster center, i.e., a geometry to which the similar geometries (cluster members) are assigned. We extracted from the MD trajectory (3M solution) 5000 geometries containing one lithium atom, one closest chlorine atom and 20 closest water molecules. These geometries were analyzed by the PIV program[2]. We were mainly interested in the description of the first solvation shell, and therefore we used the PIV focused on interatomic distances in the range of 2.1 to 6 Å. The clustering was done by Daura's algorithm[3] with a cutoff of 1.0 (distance between analyzed frames). The clustering algorithm sorted the geometries into 14 clusters.

The results of the PIV analysis are represented in Supplementary Figure S2 for the solvent-shared (SShP) (A) and contact (CP) (B) ion pair configurations. As seen, the positions of heavy atoms in the former case agree well with the distances present in our SShP cluster model. For the case of CP case, we found a difference of 0.2 Å between lithium and distant water molecules. The distances between lithium and its nearest water molecules in the MD simulations and in our CP cluster model agree well.

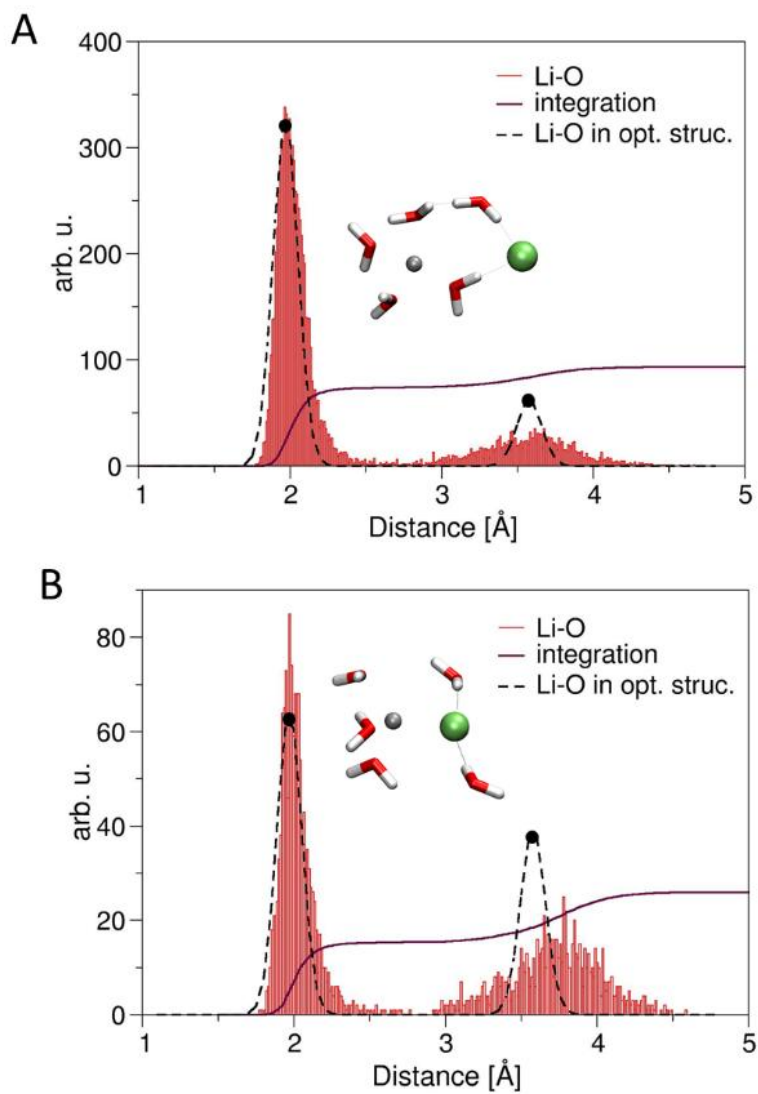

**Supplementary Figure S2 A**: Clusters with the highest number of members contain lithium and chlorine pair at an average distance of 4.3 Å which corresponds to the SShP cluster model used in our computations. The geometry of a representative cluster center containing lithium, chlorine and five closest water molecules is depicted in the inset. The distances between the central lithium atom and the five closest oxygen atoms for all members of the cluster are shown as a histogram. The black points show the distances between lithium and oxygen atoms in our SShP model. In this model, four water molecules are positioned at the distance of 1.95 Å and one water

molecule is positioned at the distance of 3.53 Å from the lithium atom. The dashed line shows Gaussian distributions of the Li⋯O distances with an artificial broadening. The ratio of the peak intensities corresponds to 4:1. The integration of the MD histogram (shown as a violet line) produces the ratio of 3.78:1 between the first peak (at around 2.0 Å) and the second peak (at around 3.6 Å). **B**: The same analysis performed for clusters with lithium and chlorine atoms separated by an average distance of 1.9 Å (correspond to the CP model). The geometry of a representative cluster center containing lithium, chlorine and five closest water molecules is depicted in the inset. In the CP cluster model used in our computations there are three water molecules closest to lithium and two water molecules at a larger distance. The corresponding values are indicated by dots. The dashed line shows Gaussian peaks with the intensity ratio of 3:2. The integration of the histogram gives the ratio 3:2.1 between the first (2.0 Å) and second peak (3.8 Å).

**Impact of the local solvent structure on ETMD spectra**

ETMD is a charge transfer process and hence involves predominantly nearest neighbors. Besides being sensitive to ion pairing, ETMD processes depend also on the geometry of the local solvent structure. This is seen from Supplementary Figures S3 and S4 which demonstrate the effects of water orientation and lithium-water distance on ETMD spectra, respectively. The overall effect is not large, yet one can observe certain trends. The black curves in panels (a)-(c) of Figure S3 are the computed ETMD spectra taken from panels (d)-(f) of Figure 2 of the paper, respectively. They correspond to the SSP, SShP and CP cluster models of $C_1$ symmetry (see Computations for geometry optimization details). By keeping the ion-ion and ion-water distances, the cluster geometries have been re-optimized. As a constrain, we imposed $C_s$ symmetry of the re-optimized structures. The geometry re-optimization led to changes in the relative positions and orientations of water molecules, in particular the hydrogen atoms of the water molecule located above the lithium (see the insets in Figure S3) lie now in the plane of inversion. The changes in the orientations of water molecules are most reflected in the spectra.

Let us consider the SShP model where the spectral changes are most pronounced (panel (b) of Figure S3). Due to the fact that the $1b_1$ orbital of the aforementioned water molecule has now a smaller overlap with the lithium ion, the low-energy region of the main ETMD(3)$_{W,W}$ peak at 26 eV ascribed to the ETMD(3)$_{1b1,3a1}$ processes has reduced its intensity. In contrast, the high-energy side of the main peak at 33 eV which is attributed to the ETMD(3)$_{1b2,3a1}$ processes has gained intensity such that even a shoulder has appeared. This is because in the re-oriented water molecule, the $1b_2$ and $3a_1$ orbitals start to better overlap with lithium.

Water rotations have also an effect on the ETMD(2)$_W$ spectral feature. In the ETMD spectrum of the SShP cluster model of $C_1$ symmetry, this feature has two peaks ascribed to the ETMD(2)$_{1b1,3a1}$ (36.5 eV) and ETMD(2)$_{3a1,3a1}$ (39 eV) processes, respectively (note that the indicated orbitals belong to the same molecule). In the spectrum of the $C_s$ system, the former peak disappears, while the latter gains more intensity.

By the example of the SShP cluster model of $C_s$ symmetry we also checked the effect of the lithium-water distance on the ETMD spectral shape (see Figure S4). Since all ETMD processes involve water, their efficiencies depend on the lithium-water separations. With decreasing lithium-water distances, ETMD becomes more efficient and the intensities of all peaks thus increase. The largest intensity growth is, however, observed for the main ETMD(3)$_{W,W}$ peak since two water molecules are involved in the respective processes.

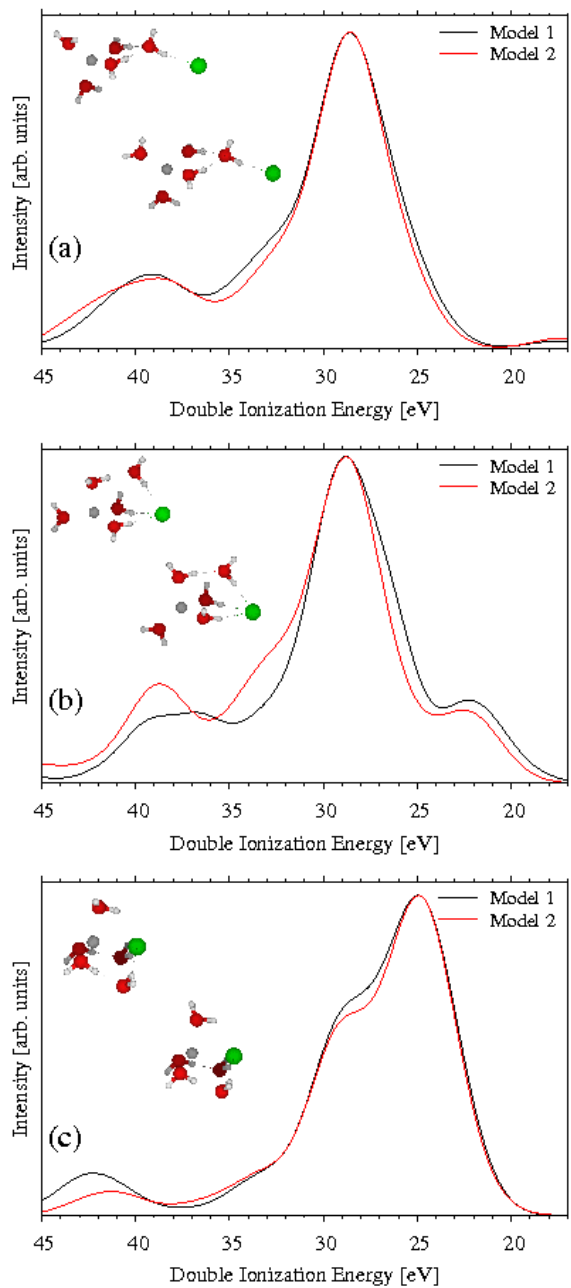

**Supplementary Figure S3** Computed ETMD spectra for three types of cluster models: (a) the SSP model, (b) the SShP model and (c) the CP model. Two different structures were considered in each case: an optimized structure of $C_1$ symmetry (black curve, the upper snapshot in the inset) and a structure of $C_s$ symmetry (red curve, the bottom snapshot in the inset). See text for geometrical differences between all models.

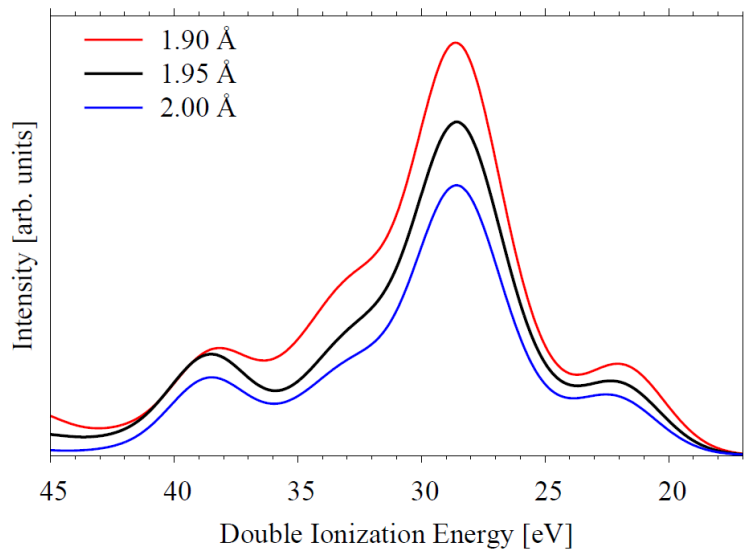

**Supplementary Figure S4** Dependence of the computed ETMD spectrum on the separation between the lithium ion and its closest (four) water molecules. The black curve corresponds to the SShP cluster model of $C_s$ symmetry (shown also as the red curve in panel (b) of Figure S3). The distances between lithium and all water molecules change simultaneously.